\documentclass[aps, preprintnumbers, nofootinbib, floatfix, amsmath, amssymb, runinaddress, tightenlines, 11pt]{revtex4}
\pdfoutput=1

\usepackage{amssymb,amsmath,amsfonts,amsbsy,slashed,graphicx}

\newcommand{\eq}[2]{\begin{equation}\label{#1}#2 \end{equation}}

\dimen\footins=1truein

\usepackage{xcolor}

\newcommand{\mpl}{M_{\rm pl}}

\newcommand{\calR}{{\cal R}}

\newcommand{\calN}{{\cal N}}
\newcommand{\calO}{{\cal O}}

\def\bea{\begin{eqnarray}}
\def\eea{\end{eqnarray}}
\def\be{\begin{equation}}
\def\ee{\end{equation}}
\def\ba{\begin{array}}
\def\ea{\end{array}}

\begin{document}

\preprint{CERN-PH-TH-2015-179}
\author{Subodh P. Patil$^{1}$}
\email{subodh.patil@unige.ch}
\author{Pedro Schwaller$^{2}$}
\email{pedro.schwaller@cern.ch}
\affiliation{1) Dept. of Theoretical Physics, University of Geneva, 24 Quai Ansermet, CH-1211 Geneva-4, Switzerland\\}
\affiliation{\\2) Theory Division, PH-TH Case C01600, CERN, CH-1211 Geneva-23, Switzerland\\}
\date{\today}

\title{Relaxing the Electroweak Scale: the Role of Broken dS Symmetry}

\begin{abstract}
Recently, a novel mechanism to address the hierarchy problem has been proposed \cite{Graham:2015cka}, where the hierarchy between weak scale physics and any putative `cutoff' $M$ is translated into a parametrically large field excursion for the so-called relaxion field, driving the Higgs mass to values much less than $M$ through cosmological dynamics. In its simplest incarnation, the relaxion mechanism requires nothing beyond the standard model other than an axion (the relaxion field) and an inflaton.  In this note, we critically re-examine the requirements for successfully realizing the relaxion mechanism and point out that parametrically larger field excursions can be obtained for a given number of e-folds by simply requiring that the background break exact de Sitter invariance. We discuss several corollaries of this observation, including the interplay between the upper bound on the scale $M$ and the order parameter $\epsilon$ associated with the breaking of dS symmetry, and entertain the possibility that the relaxion could play the role of a curvaton. We find that a successful realization of the mechanism is possible with as few as $\calO (10^3)$ e-foldings, albeit with a reduced cutoff $M \sim 10^6$ GeV for a dark QCD axion and outline a minimal scenario that can be made consistent with CMB observations. 
\end{abstract}

\maketitle

\section{Introductory remarks}
Interacting scalar fields are notoriously sensitive to the heaviest particles they couple to. Since the discovery of the Higgs boson with a relatively light mass of 125 GeV \cite{Aad:2012tfa, Chatrchyan:2012ufa}, explanations that dynamically account for the apparent hierarchy between the electroweak (EW) scale and any new physics that is presumed to complete the weak sector of the standard model\footnote{Such as low energy supersymmetry and large/ warped extra dimensions for completions that incorporate gravity, and composite models for completions which become relevant at lower energies.} appear to be in tension with atomic physics and collider constraints excepting rather tuned regions of parameter space. This has led to anthropic arguments gaining currency as a plausible alternative, although as of yet no convincing solution to the problem of how to define probabilities for observers and observables is available. Evidently, novel solutions to the hierarchy problem that circumvent current low energy constraints need no further justification. 

Recently, the authors~\cite{Graham:2015cka} have proposed a mechanism where the hierarchy between weak scale physics and the new physics scale $M$ is paraphrased into requiring a parametrically large field excursion for a field that couples to the Higgs.\footnote{For precursors in this direction, see~\cite{Abbott:1984qf,Dvali:2003br,Dvali:2004tma}.}
In order to keep any new hierarchies introduced by this new sector to be technically natural~\cite{Wilson:1973jj,'tHooft:1979bh}, an obvious choice would be for this field to be axion-like, hence a \textit{relaxion}.
The potential for the relaxion $\phi$ coupled to the singlet component of the Higgs $h := (H^\dag H)^{1/2}$ is given by
\eq{v0}{V(\phi, h) = \left(-M^2 + g\phi\right)h^2 + gM^2\phi + ... + \Lambda^4(\langle h\rangle)\,{\rm cos}\left(\phi/f\right), }
where the ellipses denote higher order terms\footnote{The expansion is arranged in this way because it is technically natural for $g$  to be small, since a discrete shift symmetry is recovered in the limit where $g$ vanishes. We will address the relevance of the higher order terms in the next section.} in $g\phi$, and $\langle h \rangle$ is the ($\phi$ dependent) vacuum expectation value of the Higgs. We presume the relaxion to begin at very large field values $\phi \gg M^2/g$ wherein the Higgs has a naturally large (and positive) mass squared. The relaxion evolves under the influence of the background cosmology which has to last long enough for $\phi$ to scan a sufficient range in field space to eventually break EW symmetry at $\phi \sim M^2/g$. Primordial inflation provides a natural context for this evolution to take place. As soon as the relaxion expectation value drops below $\phi = M^2/g$, the Higgs starts to acquire a non-zero expectation value and a periodic potential for $\phi$ is generated by instanton effects whose scale in the EW vacuum is set by
\eq{}{\Lambda^4 \sim f_\pi^2 m_\pi^2,}
where $f_\pi$ is the (non-perturbatively generated) pion decay constant and $m_\pi$ is the pion mass. Since $m_\pi^2$ grows linearly with the quark masses, this term grows in proportion to $\langle h\rangle$.  
\begin{figure}[t]
\begin{center}
\includegraphics[width=0.5\textwidth]{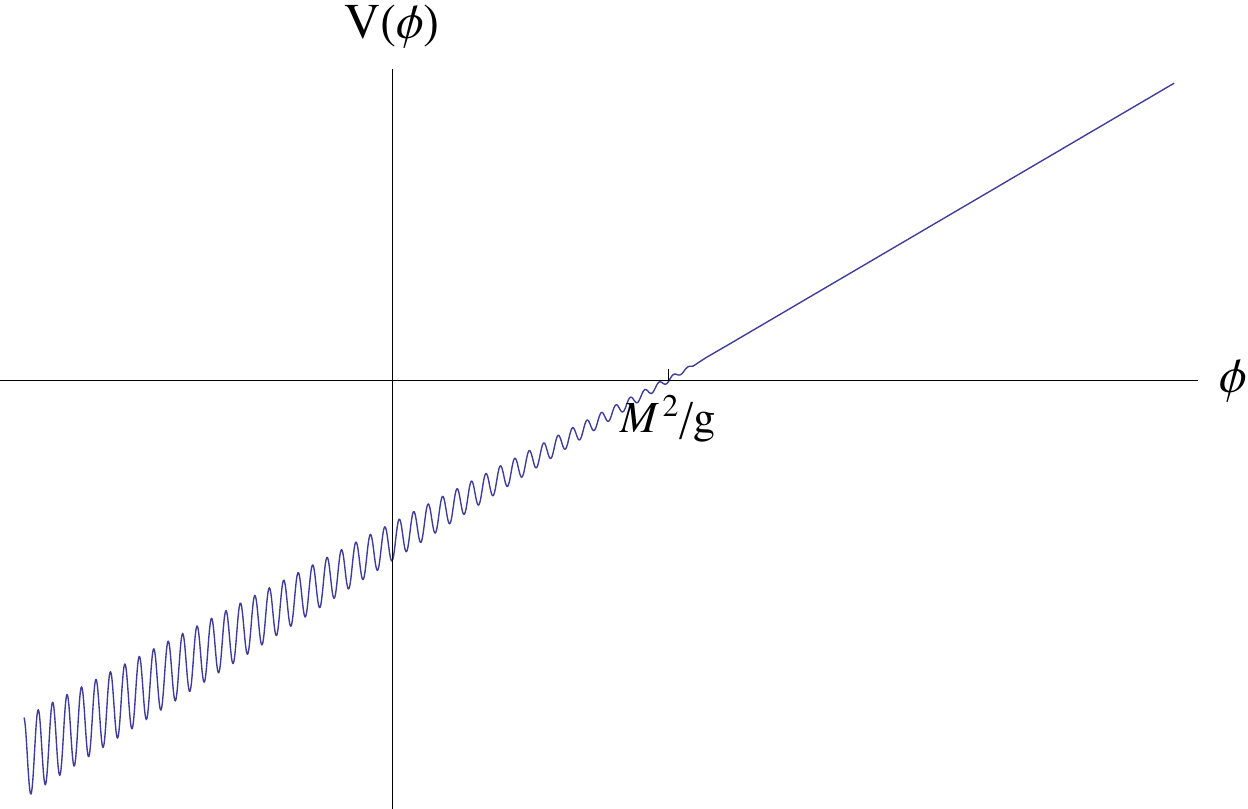}
\caption{Schematic behaviour of the relaxion potential above and below $\phi = M^2/g$. \label{potential}}
\end{center}
\end{figure}
Under the approximation that $\phi$ is slow rolling, it will get trapped in a local minimum once the barriers induced by the instanton potential are large enough to compensate the slope of the potential, which occurs when 
\eq{stop}{ \Lambda^4 \sim gM^2f.} 
Parameterising the prefactor of the periodic potential as $\Lambda^4(\langle h \rangle) = \Lambda^4 \langle h \rangle/v$, where $v = 246$~GeV, it follows that 
\eq{stoproll}{\frac{g M^2 f}{f_\pi^2 m_\pi^2} = \frac{\langle h \rangle}{v}\,.}
Since small values of $g$ are technically natural, $\frac{\langle h \rangle}{v}$ of order one can be obtained for very large values of the cut-off $M\gg v$, by adjusting $g$ accordingly. 

Thus far we have taken the cosmological history of the model for granted. Obviously, one could arrange initial conditions for $\phi$ such that it starts right above the region where the square of the Higgs mass changes sign. However in that case one would merely be trading the fine-tuning of the weak scale for a fine-tuning of initial conditions. For this mechanism to naturally explain the weak scale, one has to instead demand that the field has enough time to settle into the minimum from virtually any initial value $\phi_0$ which satisfies $-M^2 + g \phi_0 > 0$, in addition to requiring that the temperature be below the scale $\Lambda$  to guarantee that the instanton potential appears as the Higgs acquires a non-zero expectation value. Consequently, naturalness forces us to require that the relaxion undergo a minimum field excursion of $\Delta \phi \sim M^2/g$ over the course of early universe evolution. In~\cite{Graham:2015cka} (henceforth GKR) the authors implicitly considered the de Sitter (dS) limit of an inflationary background wherein $H(t) \equiv H_I$. In this case the relaxion rapidly settles onto the attractor solution
\begin{align}
	\phi(t) & = \phi_0 - \frac{gM^2}{3 H_I} (t - t_0)\,\to \Delta\phi = - \frac{gM^2}{3 H_I^2}\Delta \calN
\end{align}
where the latter follows from the fact that the number of e-folds $\Delta \calN := {\rm log}(a/a_0)$ goes as as $\Delta \calN = H_I(t-t_0)$ on a dS background. Therefore a field excursion $|\Delta \phi| \gtrsim M^2/g$ requires
\eq{Nmin}{\Delta \calN \gtrsim \frac{ 3H_I^2}{g^2} \,,} 
where we recall from (\ref{v0}) that $g$ has dimensions of mass. Since $\Delta \calN$ is bounded from below by a ratio of two largely independent scales, it should be immediately clear that $\calN$ can vastly exceed the minimum number of e-folds required to solve the horizon problem, especially considering $g$ is an a priori parametrically small quantity. %\sub{In this manner, we see that the hierarchy in energy that is the EW hierarchy problem has been translated into field excursion that is scanned by the relaxion over cosmological time scales.} 
Among the more phenomenologically viable models considered in \cite{Graham:2015cka}~anywhere between $10^{37}$ and $10^{67}$ e-folds of inflation was required\footnote{A simplified way to see this is as follows: The stopping condition for the relaxion can be written as $g/v \sim (v/M)^3$, using $\Lambda \sim v$ and $f=M$. Furthermore $H_I \geq M^2/M_{\rm pl}$ so that the inflaton dominates over the relaxion energy density. Combining these constraints in (\ref{Nmin}) one finds $\Delta \calN \gtrsim 10^{-32} (M/v)^{10}$. Taking $M\sim 10^9$~GeV then for example one finds $\Delta \calN \gtrsim 10^{38}$. }, 
admittedly a tall ask for any inflationary sector and most likely reintroducing fine tuning issues into that sector\footnote{One of the constraining requirements on inflation in the context of the relaxion mechanism is that it does not permit eternal inflation (so as to preclude the possibility of reintroducing anthropic arguments). From a model builders perspective, this is in tension with obtaining such a large number of e-folds, since models that typically do so tend to be of the chaotic variety in an eternally inflating regime.}. 

In the following section we will review the requirements of successfully realizing the relaxion mechanism and in particular, we will critically re-examine the role played by the symmetries of the background. The main result is that breaking exact de Sitter symmetry parametrically reduces the required number of e-foldings for the relaxion to undergo the requisite field excursion, greatly alleviating constraints on inflationary model building. In situations where the order parameter\footnote{Locally, de Sitter space possesses a time-like Killing vector. Inflationary cosmology is the result of spontaneously breaking this symmetry by a slow rolling background, and the curvature perturbation on constant inflaton slices relates to the Goldstone mode that non-linearly realizes this symmetry \cite{Cheung:2007st, Low:2001bw}. The slow roll parameter $\epsilon$ is an order parameter \cite{Chluba:2015bqa} in the precise sense that the action for the curvature perturbation is an expansion in powers of $\epsilon$ and its derivatives, which becomes trivial in the limit that the symmetry is restored.} associated with breaking dS symmetry -- $\epsilon := -\dot H/H^2$ -- is non-trivial, we will show that
\eq{nbound1}{ \Delta\calN \gtrsim {\rm log}\left(1 + 2\epsilon_0\frac{3 H_0^2}{g^2}\right)^{\frac{1}{2\epsilon_0}},}
where $H_0 = H(t_0)$. In the limit $2\epsilon_0 \to 0$ the quantity in the parentheses is the very definition of $e^{3H_0^2/g^2}$, cf. (\ref{Nmin}). An inflating background for which $\epsilon_0 \sim \calO(10^{-2})$ implies $\Delta \calN \sim \mathcal O(10^3)$ for a dark QCD relaxion for example, a far more manageable proposition than requiring $10^{37}$ e-folds of exactly de Sitter inflation\footnote{Presumably, the reference \cite{Graham:2015cka} did not intend this limit, yet in neglecting corrections to background quantities over many Hubble times one implicitly works in such a limit.}. 

The source of the parametric enhancement in field excursion once exact dS symmetry is broken can be understood from the attractor solution for the relaxion, which we will shortly derive as 
\eq{exc}{\frac{d\phi}{d\calN} = -\frac{gM^2}{3 H_0^2}e^{2\epsilon_0 \calN + ...},}
where the ellipses denote terms in the exponent that grow with the time variation of $\epsilon$ (\textit{serving only to enhance the field excursion per e-fold} for backgrounds where $\epsilon$ monotonically increases, as is always the case during slow roll inflation). Evidently, parametrically more field excursion occurs on backgrounds that break dS symmetry with a larger value of $\epsilon_0$. This parametric enhancement can be understood intuitively from the fact that de Sitter space is a maximally symmetric spacetime upon which freely propagating fields will have homogeneous attractor solutions that are uniformly sourced by the background geometry. Breaking this maximal symmetry results in additional source terms from the background on top of that already present in the effective potential of the field, enhancing the amount of field excursion per e-folding. Although this additional sourcing is nominally a small local correction, it integrates over many e-folds to substantially correct the bound (\ref{Nmin}) into (\ref{nbound1}). Another way of intuitively understanding this parametric enhancement is to realize that there is an obvious limit in which an infinite amount of field excursion is obtained per e-fold -- when the background asymptotes to Minkowski space. Breaking dS symmetry merely allows one to interpolate between these maximally symmetric limiting cases, where the field excursion per efold (for a fixed unit time scale $g^{-1}$) is enhanced as per (\ref{exc}).

In the subsequent sections, we first re-examine the model requirements for getting enough relaxion field excursion on a background that breaks dS symmetry. We then rederive the various constraints imposed by requiring a successful realization of the relaxion mechanism, finding an interesting interplay between the order parameter $\epsilon$ and maximum possible cutoff $M$. Furthermore, we find that accounting for corrections to the Hubble factor during inflation obliges us to discard the inflaton as the source of the curvature perturbations for regions of parameter space where $\epsilon_0 \gtrsim \calO(10^{-26})$, necessitating some other method to generate them. The curvaton mechanism offers a plausibly (and technically) natural mechanism, for which we will audition the relaxion itself for the role. We conclude by discussing various possible generalizations and cosmological aspects of the relaxion mechanism. 

\section{Cosmological relaxation revisited}

The relaxion mechanism depends on the background cosmology in several crucial respects, and so it pays to re-examine how one can satisfy all that is required of the relaxion field in as general terms as possible. The primary requirement is for enough field excursion to have been executed over the course of early universe evolution. An anxiety expressed in \cite{Graham:2015cka} is that this apparently requires a very large number of e-folds  (at least $\sim 10^{37}$ depending on the model construction). In what follows, we demonstrate that this is an artefact of deriving various bounds without accounting for finite changes in the Hubble factor during inflation. Incorporating these corrections greatly ameliorates the situation in that $\phi$ executes a parametrically larger excursion on a quasi de Sitter background.

\subsection{Field excursions and the breaking of dS symmetry}

We recall that the most general potential for the relaxion $\phi$ coupled to the singlet component of the Higgs $h$ is given (above the EW phase transition) by
\eq{ipot}{V(\phi, h) = \left(-M^2 + g \phi\right)h^2 + c_1 gM^2\phi + c_2 g^2\phi^2+... }
where the ellipses denote higher order terms in $g\phi$. Before the Higgs acquires a vev, the equation of motion for the spatial zero mode of the relaxion on an FRW background is given by
\eq{full}{\ddot\phi + 3H\dot\phi + 2c_2 g^2\phi = -gc_1 M^2\,.}
It is only in the regime where $\phi \ll M^2/g$ that one can neglect the effective mass term in the above relative to the constant driving term. Given that $\phi \gtrsim M^2/g$ for the majority of the evolution of the relaxion, we clearly cannot avoid accounting for the former. Furthermore, we note that under the field redefinition
\eq{fred}{\phi = \bar\phi + \mu}
and the choice $\mu = -c_1M^2/(2c_2 g)$, we can redefine away the linear term completely to arrive at the potential
\eq{}{V(\phi, h) = \left(-M^2\left[1 + \frac{c_1}{2c_2}\right] + g \bar\phi\right)h^2  + c_2 g^2\bar\phi^2 - \frac{c_1^2}{4c_2}M^4 +... }
highlighting the manner in which the relaxion mechanism now intertwines the hierarchy problem with the bare cosmological constant problem\footnote{Since we require the combined contributions to the cosmological constant including that which comes from the inflaton sector to vanish (or be rendered gravitationally inert e.g. \cite{Dvali:2002pe, Dvali:2007kt, Aghababaie:2003wz, Burgess:2011va}) close to the EW vacuum.}. In the  following however, we simply posit the vanishing of $c_2$ and set $c_1 =1$ for economy of discussion since we are only interested in deriving order of magnitude bounds. We demonstrate the persistence of the conclusions of this section accounting for higher order corrections to the potential in the appendix.

Writing the FRW scale factor as $a(t):= e^{\calN(t)}$ and reparametrizing time so that $\calN$ is our new clock through the relation $d\calN = H dt$, the equation of motion for a minimally coupled (test\footnote{We shall later enforce the consistency of this assumption by requiring that the relaxion energy density be sub-dominant throughout.}) scalar field on an arbitrary FRW background with the potential $V(\phi)$ is given by
\eq{eom}{\phi'' + \left(3-\epsilon\right)\phi' + \frac{V_{,\phi}}{H^2} = 0\,,}
with primes denoting derivatives with respect to $N$ and where by definition
\eq{epsint}{\epsilon(\calN) = -H'/H.} 
The above can readily be integrated as
\eq{hsol}{H(\calN) = e^{-\int^\calN_0\epsilon(\calN')\, d\calN' }H_0}
so that (\ref{eom}) becomes
\eq{}{\phi'' + \left(3-\epsilon\right)\phi' + \frac{V_{,\phi}}{H_0^2}\,e^{2 \int_0^\calN\, \epsilon(\calN)\,d \calN'} = 0.} 
We stress that the above expression is completely general. Given (\ref{ipot}) with $c_2 = 0, c_1 = 1$ we find that on an \textit{arbitrary} FRW background, the relaxion evolves according to
\eq{eom2}{ \phi'' + \left(3-\epsilon\right)\phi' + \frac{M^2 g}{H_0^2}\,e^{2 \int_0^\calN\, \epsilon(\calN)\,d \calN'} = 0.}
Given $\epsilon$ as a function of $\calN$ we can readily integrate the above. The exact solution for the attractor \textit{on an arbitrary background} is given by
\eq{arb}{\frac{d\phi}{d\calN} = -\frac{M^2 g}{H_0^2}e^{-3\calN} e^{\int^\calN_0\epsilon(\calN') d\calN'}\int^\calN_0 e^{3 z} e^{\int^z_0\epsilon(\calN') d\calN'} dz}
so that in the case where $\epsilon$ is a (not necessarily small) constant $\epsilon_0$, we can immediately conclude that
\eq{}{\frac{d\phi}{d\calN} = -\frac{M^2g\,e^{2\calN\epsilon_0}}{H_0^2(3 + \epsilon_0)},}
which the limit where $\epsilon_0 \ll 1$ corresponds to the previously advertised expression (\ref{exc}). Furthermore, we can integrate the above to find
\eq{phisol}{\phi = \phi_0 + \frac{M^2g}{2\epsilon_0 H_0^2(3+\epsilon_0)}\left[1 - e^{2\calN\epsilon_0}\right]}
The expression (\ref{arb}) corresponds to the attractor solution, neglecting the irrelevant (exponentially) decaying mode. From the above, we can immediately justify our first claim-- that backgrounds that deviate from the de Sitter limit ($\epsilon_0\to 0$) effect parametrically larger field excursions:  
\eq{phibound}{\left | \frac{d\phi}{d\calN} \right| \gtrsim \frac{M^2g}{3 H_0^2}\,e^{2\calN\epsilon_0},}
where the inequality arises from (\ref{arb}) so that as claimed, incorporating any time dependence in $\epsilon$ (recalling that it is a positive, monotonically growing parameter during slow roll inflation) serves only to further enhance the field excursion per e-fold (\ref{phibound}).

\subsection{Cosmological bounds on the relaxion}
Naturalness requirements imply that the relaxion mechanism cannot resort to fine tuning of the initial conditions. In practice this means that we must ensure that the cosmological evolution be such that the relaxion scans a sufficiently large field range as to make its initial displacement immaterial\footnote{On an expanding background, memory of the initial velocities are rapidly lost as the solution settles onto an attractor trajectory within a few e-folds.}. Therefore from (\ref{ipot}) we infer that we must require the total field excursion be such that $|\Delta\phi| = \phi_0 - \phi \gtrsim M^2/g$, so that the minimum number of e-folds required is given by
\eq{nbound2}{ \Delta \calN_{\rm min} \gtrsim {\rm log}\left(1 + 2\epsilon_0\frac{3 H_0^2}{g^2}\right)^{\frac{1}{2\epsilon_0}} \,}
which follows directly from (\ref{phisol}). In the limit $2\epsilon_0 \to 0$ the right hand side of the above uniformly converges to $\Delta\calN \gtrsim 3H_0^2/g^2$ which was the original (formidable) bound uncovered in \cite{Graham:2015cka}. Here and in the following we will assume that $\epsilon=\epsilon_0$ is constant. Incorporating the time variation of $\epsilon$ into our analysis only enhances (\ref{phibound}), such that bounds obtained using $\epsilon= \epsilon_0$ are conservative. 
Very clearly, we see that allowing for the fact that the background geometry breaks maximal symmetry logarithmically alleviates the number of e-folds required to effect the requisite field excursion. For $\epsilon_0 \gg g^2/H_0^2$, the required number of e-folds simplifies to 
\eq{nbound3}{\Delta \calN_{\rm min} \gtrsim \frac{1}{2 \epsilon_0} \log \left(\frac{3 H_0^2}{g^2} \right), }
where in neglecting terms of order $\log(\epsilon_0)/\epsilon_0$, (\ref{nbound3}) is a slightly stronger bound than~(\ref{nbound2}). 

One can also recast the constraint on the minimal number of e-folds required into an upper bound on the Hubble scale $H_f$ at the end of inflation. From the solution (\ref{hsol}) we find that
\eq{hfn}{ H_f = H_0e^{-\epsilon_0\calN}\,, }
where $\calN$ is now the total number of e-folds. This implies via (\ref{nbound2}) that
\eq{hfmax}{ H_f \lesssim \frac{H_0}{\left(1 + 6\epsilon_0 \frac{H_0^2}{g^2}\right)^{1/2}} \approx \frac{g}{\sqrt{6 \epsilon_0}},}
where the last expression again is valid for $\epsilon_0 \gg g^2/H_0^2$. 

The attractor solution for the relaxion (\ref{phisol}) was derived under the assumption that $\phi$ could be treated as a test field over the background sourced by inflation. For this to consistently be true throughout the cosmological evolution, its energy density $\rho_\phi$ should be negligible compared to the energy density in the inflation sector, which is given by $\rho_{\rm inf} = 3 H^2 M_{\rm pl}^2$. Using $\rho_\phi = H^2\phi'^2/2 + M^2g\phi$, the result~(\ref{phisol}) and $H(\calN) = H_0 e^{-\epsilon_0\calN}$ we find
\eq{rhophi}{\rho_\phi = \frac{g^2M^4}{2H_0^2(3 + \epsilon_0)^2}\left[1 + \frac{3}{\epsilon_0}\left(1 - e^{2\calN\epsilon_0}\right)\right] + \phi_0M^2 g\,. }
Requiring that $\phi_0 \gtrsim 2M^2/g$, it follows that $\rho_\phi \lesssim 3H^2 \mpl^2$ will always be true if \footnote{This bound follows from the resulting quadratic inequality for $e^{2\calN\epsilon}$, which will always be satisfied provided the associated quadratic form have no real roots.}
\eq{rhophi2}{ M^2\left(\frac{g^2}{2H_0^2\epsilon_0(3 + \epsilon_0)} + 2  \right) \lesssim  \frac{6 g \mpl}{\sqrt{2\epsilon_0}(3 + \epsilon_0)}\,.}
The conditions (\ref{hfmax}) and (\ref{rhophi2}) provide the strongest bounds on the model. GKR further require that the field $\phi$ slow-rolls and that the evolution of $\phi$ is dominated by classical fluctuations, and indeed those constraints are also easily satisfied here. 
More precisely, $\Delta \phi_{\rm cl} > \Delta \phi_{\rm quant}$ implies $g M^2 > 3/(2\pi) H^3$. This can be satisfied by choosing $H_0$ sufficiently low, however since $H$ now drops over time, this constraint will be satisfied eventually, and furthermore it is not clear whether it is a necessary condition~\cite{Graham:2015cka}. In addition the barriers that eventually stop the rolling of $\phi$ (once the Higgs acquires a vev) will only develop if $H \leq H_0 < \Lambda$, where the stopping condition $\Lambda^4 \sim g M^2 f$ has to be satisfied\footnote{\label{caveat}Note that we only require $H \leq \Lambda$ before EW symmetry breaking, and $H_0$ could be larger than $\Lambda$ at the beginning of inflation provided it eventually drops below $\Lambda$ (cf. (\ref{hfn})). Although this is a reasonable possibility that doesn't introduce particularly severe tuning of the initial conditions, it is an extra ingredient that we avoid for the sake of conservativeness.}. 

We will now consider different limits of (\ref{rhophi2}). Note that during inflation we always have $\epsilon_0 < 1$, by definition. The two cases that we can immediately distinguish are $g^2/H_0^2 \gg \epsilon_0$, which includes the exact dS limit $\epsilon_0 = 0$, and $g^2/H_0^2 \ll \epsilon_0$, where the constraints (\ref{nbound3}) and (\ref{hfmax}) simplify (and where the requisite number of e-folds is vastly reduced). 

Consider first the case $g^2/H_0^2 \ll \epsilon_0$ which permits us to ignore the first term in the left hand side of (\ref{rhophi2}). We immediately find
\eq{}{	M^2 \lesssim \frac{3}{\sqrt{2}(3+\epsilon_0)} \frac{g M_{\rm pl}}{\sqrt{\epsilon_0}}\,.
}
Utilizing the stopping condition $\Lambda^4 = g M^3$ (for simplicity we consider $M=f$), $g$ can be eliminated to obtain 
\eq{Mmax1} {M \lesssim \epsilon_0^{-1/10} \left( \Lambda^4 M_{\rm pl}\right)^{1/5}\,.}
For $\epsilon_0\sim 10^{-2}$, this implies $M \lesssim 6\mbox{ TeV}$ for $\Lambda = 1\mbox{ GeV}$ and $M \lesssim 250\mbox{ TeV}$ for $\Lambda= 100\mbox{ GeV}$, which is around the maximally allowed value for $\Lambda$ in the GKR models. Provided the axion decay constant $f$ is commensurate with $M$, the bounds just derived are accurate up to factors of order unity. The number of e-folds required for the scenario to work in this context is then 
\eq{ngtr} { \Delta \calN \gtrsim 50 \log \left( \frac{M^6}{\Lambda^6} \right) \approx 3\times 10^3\,,}
using $H_0 = \Lambda$, i.e. the conservative (cf. footnote \ref{caveat}) maximal allowed value, and $f = M$ as before. 

Now consider the opposite limit, $g^2/(H_0^2) \gg \epsilon_0$. Here the second term on the LHS of (\ref{rhophi2}) can be neglected, resulting in
\eq{Mmax2} { M^2 \lesssim \frac{12}{\sqrt{2}} \left( \sqrt{\epsilon_0} \frac{H_0}{g} \right) H_0 M_{\rm pl} \,.}
The term in brackets is less than unity in the regime we are considering. Furthermore using $H \leq H_0 \leq \Lambda$, we find $M^2 \ll \Lambda M_{\rm pl}$. This is the same constraint that GKR obtain, however here we now see that a careful adjustment of $H_0$ and $\epsilon_0$ (for a given $g$) is required to actually saturate it. Inserting $g = \Lambda^4/M^3$ we find 
\eq{Mmax3} { M \gtrsim \frac{\Lambda^4}{\sqrt{\epsilon_0} H_0^2 M_{\rm pl}} \,.}
$M$ is maximised when~(\ref{Mmax1}) and~(\ref{Mmax3}) intersect. Setting $H_0 = \Lambda$ this happens for $\epsilon_0 = (\Lambda/M_{\rm pl})^3$. 
\begin{figure}[t]
\begin{center}
\includegraphics[width=0.45\textwidth]{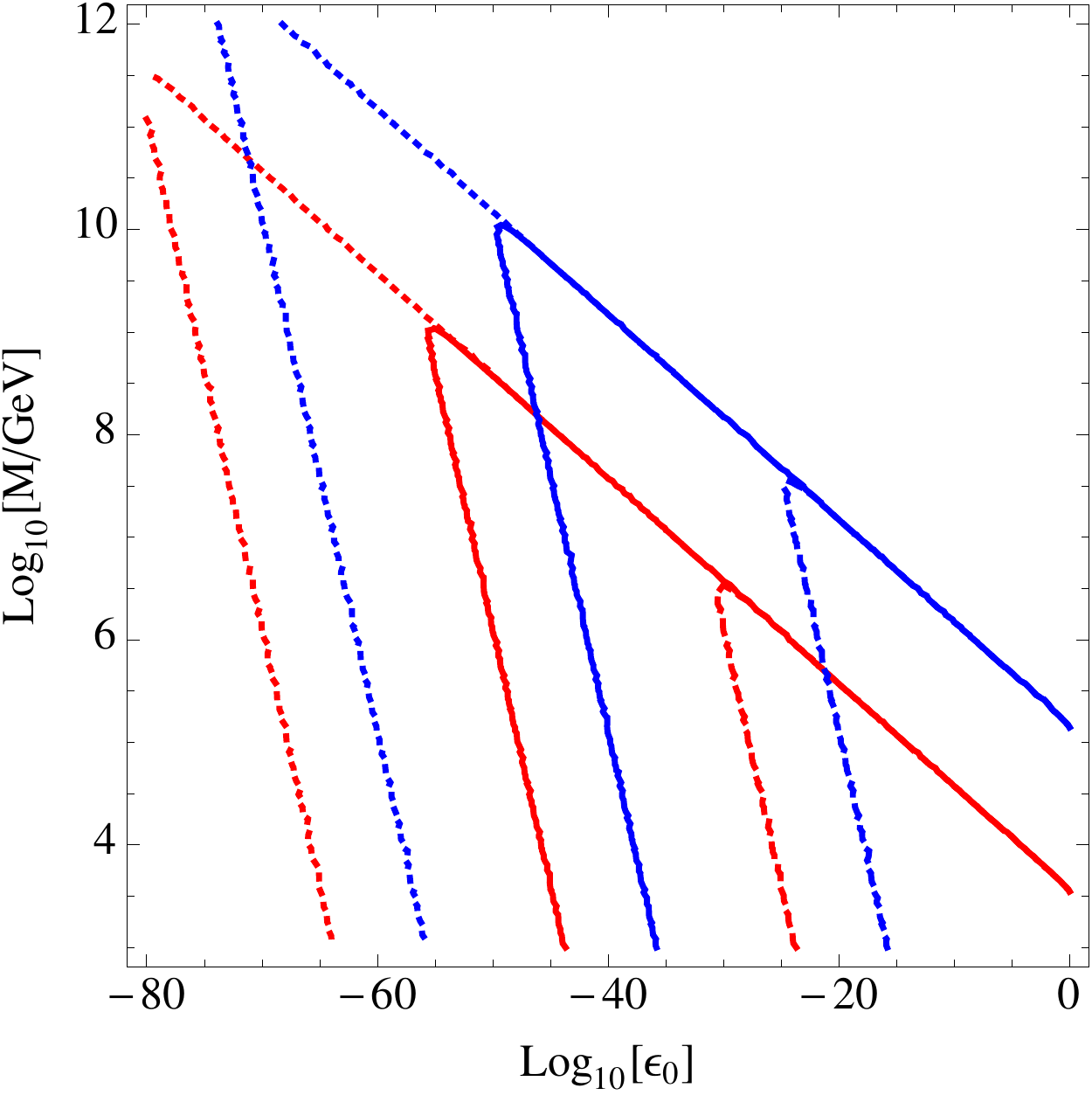}\hspace*{0.5cm}
\includegraphics[width=0.45\textwidth]{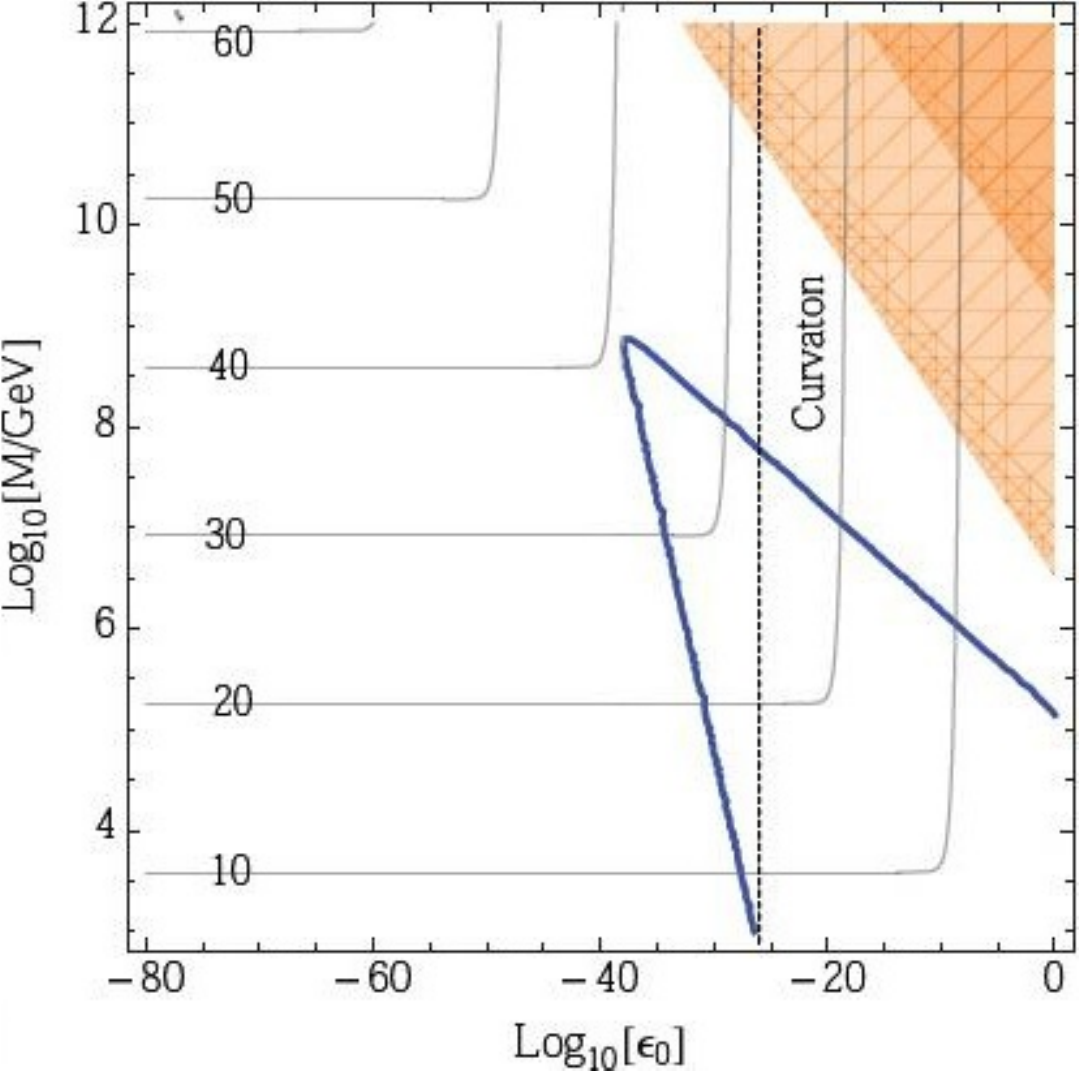}
\caption{\label{Mmax}
Left: Upper bounds on $M$ as function of $\epsilon_0$. The solid red (blue) contours correspond to $H_0 = \Lambda = 1$~GeV (100~GeV) respectively, the region consistent with~(\ref{rhophi2}) lies within the triangular shapes. Dashed (dotted) lines correspond to $H_0 = 10^{-5} \Lambda$ ($H_0 = 10^{5} \Lambda$). 
Right: Allowed region for $\Lambda = 100$~GeV, with grey lines showing $\log_{10} \calN_{\rm min}$. The light (darker) orange shaded region has $T_{RH} <\mbox{TeV}$ ($T_{RH} < 100$~MeV), and we choose $H_0$ such that $H_0 \leq \sqrt[3]{\Lambda^4/M}$ is satisfied for $M \leq 10^9$~GeV. The parameter $g$ is set everywhere by $g M^3 = \Lambda^4$. In the region to the right of the vertical dashed line, a curvaton field is needed to satisfy~(\ref{cmb}). 
}
\end{center}
\end{figure}
Fig.~\ref{Mmax} shows the constraint~(\ref{rhophi2}) in the $\epsilon_0-M$ plane for different choices of $\Lambda$ and $H_0$. The scale $M$ can be increased either by increasing $\Lambda$, which however in GKR is bounded by $\Lambda \lesssim 100$~GeV, or by increasing $H_0$, which is subject to the constraint $H_0 \lesssim \Lambda$ (unless one allows for the (relatively mild) additional requirement that the dynamics be such that $H$ eventually falls below $\Lambda$ as $\phi$ approaches $M^2/g$). In the right figure we also show $\log_{10} \calN_{\rm min}$ for the $\Lambda =100$~GeV case, which clearly highlights the correlation between the achievable cut-off and the required number of e-folds. Here we have chosen $H_0$ such that $gM^2 >3/(2\pi)  H_0^3$ is satisfied everywhere in the allowed regions, such that the field evolves classically throughout. 

\subsection{Observational Constraints}

Another constraint on the model comes from demanding successful reheating of the universe after inflation.  An upper estimate for the reheating temperature is $T_{\rm RH} \lesssim \sqrt{H_f M_{\rm pl}}$. Big bang nucleosynthesis (BBN) puts an absolute lower bound of $T_R \gtrsim 10-100$~MeV. Successful baryogenesis and dark matter production suggest $T_R \gtrsim $~TeV, however specific models can realise this also with a lower $T_{RH}$. Eqn.~(\ref{hfmax}) gives an upper bound on $H_f$, which however is always weaker than the constraints arising from~(\ref{rhophi2}). 

Important constraints also arise from the inflaton sector from the dual requirement that inflation never be in the eternally inflating regime\footnote{So as not to re-introduce a mechanism for anthropic selection, as per the requirements of \cite{Graham:2015cka}.}, and that eventually, inflation produces the observed amplitude of curvature perturbations. The first requirement implies that quantum fluctuations always be subdominant to fluctuations arising from classical rolling. That is
\eq{}{\frac{H}{2\pi} \lesssim H^{-1}|\dot\psi|\,,}
where we denote the inflaton as $\psi$. This is simply the statement that the inflaton roll further in one Hubble time than a characteristic stochastic field fluctuation (set by the temperature associated with the de Sitter horizon $T_{dS} = \frac{H}{2\pi}$ in units where the Boltzmann constant $k_B \equiv 1$ \cite{Spradlin:2001pw}). Through the equivalent definition $\epsilon = \frac{\dot\psi^2}{2\mpl^2H^2}$, the above immediately implies that throughout inflation,
\eq{eternal}{\frac{H^2}{8\pi^2\mpl^2\epsilon} \lesssim 1\,.} 
One can also recognize in the above that this is precisely the quantity that is inferred through CMB observations as \cite{Ade:2015xua}:
\eq{cmb}{\frac{H_*^2}{8\pi^2\mpl^2\epsilon_*} \simeq 2.2\times 10^{-9} \,,}
where starred quantities indicate that these are evaluated at the time some fixed comoving scale (typically taken to correspond to COBE normalization at $k = 2\times 10^{-4}$ Mpc$^{-1}$) exits the horizon. The bound (\ref{eternal}) is under most tension at the beginning of inflation. If one demands that~(\ref{cmb}) is satisfied, it implies 
\eq{efoldb}{\frac{H_0^2}{H_*^2}\lesssim \frac{10^9}{2.2}\frac{\epsilon_0}{\epsilon_*} \lesssim \frac{10^9}{2.2}.}
However at the same time it is important to note that, given the low inflation scale required by the relaxion scenario $H_0 \lesssim \Lambda \lesssim v$, the constraint (\ref{cmb}) can only be satisfied by having $\epsilon_0 \lesssim \epsilon_* \sim 10^{-30}-10^{-26}$, reintroducing the fine-tuning problem in the inflaton sector that we tried to avoid in the first place. 
Making things worse, while the amplitude of the CMB anisotropies can be fit with such a tiny $\epsilon_*$, the spectral tilt $n_s= 1 + 2\eta_* - 6\epsilon_* \approx 0.96$ \cite{Ade:2015xua} would then require $\eta_* \approx -2 \times 10^{-2}$, which is a highly non-trivial additional requirement on the inflationary sector.

An alternative mechanism to generate the CMB anisotropies is required in the parameter space to the right of the dashed vertical line in Fig. 2, if for example we want to keep $\epsilon_0\sim 10^{-2}$. 
%
%We note that according to (\ref{epsbound}), one needs to posit an alternative mechanism to generate the CMB anisotropies in the parameter space to the right of the dashed vertical line in Fig. 2. 
%
A plausiblecandidate for such a mechanism is offered by the curvaton scenario \cite{Lyth:2001nq, Lyth:2002my}, which can be engineered to be technically natural for example by having other axions play the role of the curvaton \cite{Dimopoulos:2003ii, Dimopoulos:2003az, Bozza:2003pj} (see however \cite{Mazumdar:2015pta} for concerns about axion over production during reheating). They are merely required to be light, energetically sub-dominant during inflation, uncoupled (or weakly coupled) to anything other than the radiation bath into which it subsequently decays, and have a mass $m_{\sigma}$
%\footnote{Where $\Lambda_\sigma$ and $f_\sigma$ are the non-perturbatively generated scale and the decay constant associated with this axion.} 
%$m^2_\sigma \sim \Lambda^4_\sigma/f_\sigma^2$ 
around the minimum of its potential such that $H_r \ll m_\sigma$ where $H_r$ is the Hubble factor during post-inflationary radiation domination. The spectral index generated by the curvaton is given by
\eq{curvtilt}{n_s-1 = -2\epsilon_* + 2\frac{(V_{\sigma\sigma})_*}{3H^2_*}.}
To obtain a Gaussian spectrum the curvaton has to be sufficiently displaced from its local minimum ($\sigma=0$) at the time when the observed CMB modes cross the horizon, i.e. $\sigma_* \gg H_*$, and furthermore that the curvaton energy density is sub-dominant to radiation immediately prior to its decay and that the decay occurs over a small window relative to a Hubble time \cite{Lyth:2001nq, Lyth:2002my}.
%
%where eventually obtaining a Gaussian spectrum of curvature perturbations requires the necessary (but not sufficient) condition that the curvaton be sufficiently displaced away from its local minimum at $\sigma = 0$ \cite{Lyth:2001nq, Lyth:2002my}:
%\eq{gauss}{H_* \ll \sigma_*\,,}
%where again, starred quantities indicate that these are evaluated at the time of horizon crossing of the observed CMB modes \textit{during inflation}. Gaussianity further requires that the energy density of the curvaton be sub-dominant to radiation immediately prior to its decay and that the decay occurs over a small window relative to a Hubble time \cite{Lyth:2001nq, Lyth:2002my}. 
The amplitude of the power spectrum of the curvature perturbations sourced by the curvaton in the regime where $H_* \ll \sigma_*$ is given by
\eq{radcurv}{\Delta_\calR \approx \kappa^2\frac{H^2_*}{\pi^2\sigma_*^2}  \simeq 2.2\times 10^{-9}\,,}
where $\kappa$ is the fraction of the curvaton energy density relative to radiation density just before it decays in the radiation dominated epoch \cite{Lyth:2001nq, Lyth:2002my}. We wish to emphasize in the context of (\ref{curvtilt}) that in general $\epsilon_* \geq \epsilon_0$ and that the precise predictions of the curvaton scenario supplemented with the relaxion mechanism will also depend on the details of the inflaton sector. However it should be clear that provided inflation can also be arranged to be technically natural, it should be possible to supplement the relaxion mechanism with an additional axionic curvaton sector preserving naturalness and simultaneously fitting to CMB observations. Taken together, this gives a minimal realization of the relaxion mechanism which is in agreement with current cosmological data. The even more minimal possibility that the relaxion itself plays the role of the curvaton is explored in the appendix.

We conclude this particular subsection by noting that thus far, we have only considered inflating backgrounds, i.e. $\epsilon \ll 1$. Given that our analysis was completely general, one could have also asked whether $\phi$ can roll into its minimum during ordinary radiation or matter dominated expansion. Those phases are characterised by $a(t) = t^{2/3}$ for matter domination and $a(t) = t^{1/2}$ for radiation domination, corresponding to $\epsilon = 3/2$ and $\epsilon = 2$, respectively. The constraints on the parameter space in this case simply follow from~(\ref{hfmax}) and~(\ref{Mmax1}) and are $H_f \lesssim g$ and $M \lesssim (\Lambda^4 M_{\rm pl})^{1/5}$. A priori these scenarios don't seem to be much more constrained than those with $\epsilon \sim 10^{-2}$. Hence for models where $\Lambda$ can be increased above the weak scale like the CHAIN scenario~\cite{Espinosa:2015eda}, this possibility should also be explored. In this context the proposal of~\cite{Hardy:2015laa}, namely to use a hidden sector with a larger temperature as an alternative to an inflationary background for the relaxion, is noteworthy. 

\subsection{Concrete models}
In the above discussion we have not explicitly specified which realisation of the relaxion mechanism is being considered. The QCD scenario of GKR is not viable due to predicting a too large strong CP angle, and the modified version to fix this problem requires $g \sim 10^{-30}$~GeV, such that only scenarios with tiny $\epsilon_0$ will be viable, due to the combination of the reheating constraint $T_{\rm RH} \sim \sqrt{H_f M_{\rm pl}} \gtrsim T_{\rm BBN}$ with $H_f \lesssim g/\sqrt{6 \epsilon_0}$ from~(\ref{hfmax}). 

Instead the model with an additional strongly coupled dark sector is potentially viable over a larger range of values of $\epsilon_0$. There the scale $\Lambda$ is set by $\Lambda^4 = 4 \pi f_\pi^3 y \tilde{y} \langle h \rangle^2/m_L$, where $f_\pi$ is the dark sector pion decay constant, $y$ and $\tilde{y}$ are additional Yukawa couplings and $m_L$ is the mass of a new fermion which is of order of the weak scale. Taking all parameters near the weak scale, $\Lambda\approx 100$~GeV is possible.

Supplementing this scenario with a curvaton we can arrive at a minimal scenario that is compatible with observations. Choosing the curvaton mass sufficiently small we get $n_s -1 \approx 2\epsilon_*\approx 2 \epsilon_0$, which agrees with observations for $\epsilon_0 \approx 10^{-2}$, and amplitude of perturbations, can easily be matched to the observed value. This scenario allows $M$ of order few hundred TeV, while being consistent with CMB observations. A more elaborate inflationary sector seems to be required to further increase the cut-off scale, while remaining in agreement with observations. 

As mentioned above, in~\cite{Espinosa:2015eda} a variation of the relaxion mechanism, the CHAIN mechanism, was presented which works even for $\Lambda \sim M \sim 10^9\mbox{ GeV}$. This is achieved by modifying the stopping condition to $\epsilon' v^2 = g M$, where $\epsilon'$ is a new small (technically natural) parameter (and we take $f\sim M$), which is constrained by $\epsilon'^2 \lesssim v^2/M^2$. Preforming the same steps leading to~(\ref{Mmax1}), we find 
\eq{}{M \lesssim \epsilon_0^{-1/10} \left( v^4 M_{\rm pl}\right)^{1/5} \,,
}
where we have eliminated $\epsilon'$ using $\epsilon' = v^2/M^2$. Since $v^2$ is again the weak scale here, we find the same parametric behaviour as in the GKR scenario, namely that exceptionally small values of $\epsilon_0$ are required to increase the cutoff beyond $(v^4 M_{\rm pl})^{1/5} \sim 10^6$~GeV. In practice the constraints on the scenario of~\cite{Espinosa:2015eda} are probably stronger, since there is an additional scalar field which also slow rolls and for which similar constraints as for $\phi$ have to be satisfied. 

In the CHAIN scenario the criterion for the instanton potential to be present is easy to satisfy even in an ordinary radiation or matter dominated universe, only requiring $T \lesssim M$, or equivalently $H_0 \lesssim M^2/M_{\rm pl}$, since the barriers now appear at the scale $M$. Therefore it seems possible that the CHAIN relaxation happens during a radiation or matter dominated period of the early universe. If this is not followed by a period of inflation, one then faces the problem that different Hubble patches of the visible universe will have different (but similar) values of $v^2$, due to quantum spreading of the relaxion~\cite{Graham:2015cka,Espinosa:2015eda}. Any scenario where there is no inflation after the relaxion stops evolving, or where the end of inflation coincides with the relaxion settling in a local minimum, will have to address this issue.

\section{Concluding remarks}

In paraphrasing the electro-weak hierarchy into a super cut-off field excursion for the relaxion, the GKR mechanism plausibly addresses the hierarchy problem in a technically natural manner. One of the drawbacks of the original scenario was the astronomical number of e-folds apparently required of the inflationary sector (no less than $\sim 10^{37}$ e-folds depending on the model construction) in order to effect enough field excursion for the relaxion. In this note, we uncovered how incorporating corrections to background quantities over many Hubble times alleviates the required number of e-folds in situations where the background breaks dS symmetry. We observed an interesting interplay between the order parameter $\epsilon$ associated with breaking dS symmetry and the maximum cutoff $M$ attainable in a given construction. In comparison with the findings of~\cite{Graham:2015cka}, we find that one can reduce the required number of e-folds down to order $10^3$ for a dark QCD axion with $\Lambda = 100$ GeV, at the expense of having to lower the cutoff to $M\lesssim 3\times 10^5$~GeV. The cutoff increases with $\epsilon_0^{-1/10}$ while the number of e-folds scales as $1/\epsilon_0$, therefore increasing $M$ by an order of magnitude increases the required number of e-folds by a factor $10^{10}$. 

An additional price paid for lowering the number of e-folds is that the inflaton can no longer be the source of the observed density perturbations for a large region of parameter space. This situation is apparently no worse than in~\cite{Graham:2015cka} where although the amplitude of the perturbations can be fit, $n_s$ appears to be much closer to unity than allowed by data. We instead propose the curvaton mechanism as viable source of the density perturbations, which for $\epsilon_0 \sim 10^{-2}$ can be made to fit both the amplitude and spectrum of the CMB anisotropies. One preserves technical naturalness by default if the curvaton is also an axionic degree of freedom and provides a minimal framework to be in agreement with astrophysical observations. As we show in the appendix, the relaxion itself can plausibly play the role of the curvaton at the expense of introducing fine tuning of initial conditions. Devising a technically natural inflaton plus curvaton sector remains a task for future investigations.

Models with a dark QCD near or below the weak scale seem to be preferred for the minimal relaxion mechanism to work, and a question that arises immediately in this context is why the scale of the dark sector should be close to the weak scale. One possibility to dynamically connect the scales is provided by approximate infrared fixed points in the running of the gauge couplings~\cite{Bai:2013xga}. This construction requires additional states in the few TeV range, which could also serve as portals to allow entropy transfer from the dark QCD to the visible sector. Possible collider signatures in that case include displaced decays of dark mesons or glueballs~\cite{Juknevich:2009gg,Craig:2015pha,Curtin:2015fna}, emerging jets~\cite{Schwaller:2015gea}, or other typical signatures of hidden valley models~\cite{Strassler:2006im}, depending on the exact mass scales involved. Since the new states in the hidden sector can not decouple without spoiling the relaxion mechanism (unless additional relaxion fields are introduced~\cite{Espinosa:2015eda}), a careful study of the expected collider signals and the effects on electroweak precision observables would be valuable.

Another interesting aspect concerns the cosmology of these models. Besides the obvious task of constructing viable (and natural) inflationary backgrounds, the generation of the observed baryon asymmetry becomes a relevant problem now. Presumably a reheating temperature below $\Lambda$ is preferred to prevent the relaxion from evolving after the end of inflation. Cold baryogenesis~\cite{GarciaBellido:1999sv,Krauss:1999ng,Tranberg:2003gi,Servant:2014bla} is an obvious candidate mechanism for producing an asymmetry in this environment, and it would be interesting to determine whether it can be implemented without introducing new fine tuning problems. 

\acknowledgements{We wish to thank B.~Batell, C.~Burgess, A.~Delgado, P.~Fox, G.~Giudice, M.~McCullough, A.~Pomarol, I.~Sawicki, D.~Stolarski, A.~Strumia and A.~Weiler for useful and inspiring discussions, and B.~Batell, G.~Giudice and A.~Riotto for helpful comments on the manuscript. PS would also like to acknowledge support from the Munich Institute for Astro- and Particle Physics (MIAPP) of the DFG cluster of excellence "Origin and Structure of the Universe", where part of this work was completed. }

\begin{appendix}

\section{Field redefinitions and higher order corrections}

The relaxion potential as introduced in \cite{Graham:2015cka} is an expansion in powers of $(g \phi/M^2)$, where $M$ is the scale that characterizes the UV completion of the theory. Therefore, the full potential can be written as 
\eq{} { V(\phi) = M^4 \sum c_n \left( \frac{g \phi}{M^2} \right)^n\,.
}
For small field values this is clearly dominated by the linear term, however the relaxion mechanism can only be considered compatible with naturalness if we allow for field excursions $\Delta \phi \gtrsim M^2/g$, in which case all terms in the potential could become equally important. To see that the linear term accurately captures the dynamics over a field range approximating $\Delta\phi \sim M^2/g$, it is useful to expand the field around the EW symmetry breaking field value $M^2/g$ as $\phi = M^2/g + \sigma$, since we are interested in the dynamics of $\phi$ near that point. The potential for $\sigma$ then becomes
\eq{} {V(\sigma)  = c_0' + c_1' g M^2 \sigma + {\cal O}(g^2 \sigma^2) \,,
}
where now the higher order terms are negligible (since $\sigma < M^2/g$) and where $c_0'$ contributes to the bare cosmological constant and $c_1' = \sum_n n c_n$. We are therefore justified in only considering the linear term for the evolution of $\phi$ in the region we are interested in, namely near $\phi = M^2/g$. The higher order terms in the potential can be absorbed in a redefinition of $g$ and $M^2$. We note that requiring that $\sum_n n c_n$ be finite is readily conceivable if the UV completion describes weakly coupled dynamics of the putative UV degrees of freedom.

\section{The relaxion as a curvaton}

We had previously observed that for backgrounds for with $\epsilon \gtrsim \calO(10^{-34})$, the curvature perturbations have to be generated by some mechanism other than the standard inflationary one. The curvaton mechanism offers a viable candidate, and it is natural to ask whether the relaxion could play the part of the curvaton. 

As detailed in \cite{Lyth:2001nq, Lyth:2002my}, the curvaton mechanism relies upon an effectively light spectator field during inflation (which acquires a close to scale invariant spectrum by default), which begins to oscillate around the minimum of its potential during radiation domination (i.e. after reheating). In order for a scale invariant, Gaussian spectrum of curvature perturbations to be generated\footnote{This generation occurs via isocurvature perturbations (corresponding to the curvaton) sourcing curvature perturbations through the presence of non-adiabatic pressure perturbations during radiation domination. Consequently, the spectral properties of the curvaton spectrum imprint onto the curvature perturbation as the curvaton decays.} three conditions have to be satisfied: i) the Hubble factor during radiation domination be such that $H_r < m_\sigma$ so that the relaxion be able to oscillate around its minimum; ii) that as the modes that we observe in the CMB exited the horizon, $H_* \ll \sigma_* $ so that the perturbations $\delta\sigma/\sigma_*$ remain weakly coupled (i.e. Gaussian) and iii) that the curvaton be coupled to the radiative degrees of freedom at tree level so that it can subsequently decay efficiently \cite{Lyth:2001nq}. 

The last condition is trivially satisfied since the relaxion by definition couples to gauge bosons (be they in visible or dark sectors). 
The first condition neccessitates that in addition to inflation ending before the relaxion finds its minimum, that EW symmetry breaking occurs before BBN, and that necessarily for the barriers to have formed in the first place,
\eq{hrub} {T_r = \sqrt{M_{\rm pl} H_r } \lesssim \Lambda\,.}
This can be shown to impose a tuning of initial conditions for $\phi_0$ such that $\Delta \phi_0 / (M^2/g)$ is specified to percent level precision, evidently the price to pay for not having to introduce additional degrees of freedom. 
Furthermore we require that at this time $H_r \lesssim m_\sigma$ so that the relaxion can oscillate around its minimum. That this is plausible can immediately be inferred from the fact that once EW symmetry breaking has been effected, fluctuations of the relaxion around its metastable minimum (parametrized as $\phi = M^2/g + \sigma$) will have an effective mass of 
\eq{}{m^2_\sigma \sim \frac{\Lambda^4}{f^2} \sim g M\,,} 
which follows directly from (\ref{v0}) when $\langle h \rangle = v$ and the relation (\ref{stop}) presuming the natural value for the axion decay constant $f \sim M$. Given that the implied curvaton mass ranges from $m^2_\sigma = 10^{-6}$ GeV if the relaxion is the QCD axion ($\Lambda = 1$ GeV) or $m^2_\sigma = 10^{-2}$ GeV if it is a dark QCD axion ($\Lambda = 100$ GeV), we see immediately that this is far above the upper bound on $H_r$ set by (\ref{hrub}) $H_r \lesssim \Lambda^2/\mpl \sim 10^{-14} - 10^{-18}$ GeV. Hence we conclude that efficient conversion of the isocurvature perturbations is possible in the radiation dominated phase with the relaxion as the curvaton. 

The requirement that $H_*/\sigma_* \ll 1$ during inflation ensures that the subsequent curvature perturbations will be close to Gaussian \cite{Lyth:2001nq} consistently allows for the amplitude for the power spectrum (\ref{radcurv}) to be fit to observations as
\eq{radcurv2}{\Delta_\calR \approx \kappa^2\frac{H^2_*}{\pi^2\sigma_*^2}  \simeq 2.2\times 10^{-9}}
where $\kappa$ is the fraction of the curvaton energy density to radiation immediately prior to its decay. Since we require $H \lesssim \Lambda$ throughout inflation, we see that $H_* \ll \sigma_*$ if $\Lambda \ll \sigma_*$. From  (\ref{Mmax1}) we see that for $\epsilon_0 \sim 10^{-2} - 10^{-20}$ a dark QCD axion $\Lambda = 10^2$ GeV implies that $M^2/g \sim 10^{18}$ GeV, and that since $0 \leq \sigma \leq M^2/g$, the condition $H_* \leq \Lambda \ll \sigma$ is satisfied for almost all of the permissible field range. In order to ensure the correct normalization (\ref{radcurv2}) requires however that whatever the scales $H_*$ and $\sigma_*$ during inflation, $\kappa$ has to have the precise value (during radiation domination) such that (\ref{radcurv2}) is satisfied. This can only be viewed as another source of tuning in addition to the tuning of initial conditions required by satsifying (\ref{hrub}) throughout. As discussed in \cite{Toni}, this mechanism is further constrained by the fact that unless an additional mechanism halts the relaxion immediately as it encounters its first local minimum, $\sigma_*$ is expected to be at least of order $f$ since this is the field displacement in the relaxion potential equivalent to the energy displacement implicit in the formulae above. This means that since
\eq{radcurv2}{\Delta_\calR \approx \kappa^2\frac{H^2_*}{\pi^2\sigma_*^2}  \lesssim \frac{\kappa^2}{\pi^2}\frac{\Lambda^2}{f^2}\,,}
one has yet further fine tuning of initial conditions (now in the inflaton sector) such that the COBE pivot scale we see in the CMB must have exited the horizon when the right hand side above is of the right order. As discussed below (\ref{Mmax1}) for low scale cut-off models (with $\epsilon \sim 10^{-2}$) $\Lambda/f \sim 10^{-3}$ one can plausibly accomplish this with enough tuning of parameters and initial conditions, however this rapidly becomes impossible for models with very high cut-offs (i.e. very close to dS, cf. (\ref{Mmax1})).

\end{appendix}


\begin{thebibliography}{99}

%\cite{Graham:2015cka}
\bibitem{Graham:2015cka} 
  P.~W.~Graham, D.~E.~Kaplan and S.~Rajendran,
  ``Cosmological Relaxation of the Electroweak Scale,''
  arXiv:1504.07551 [hep-ph].
  %%CITATION = ARXIV:1504.07551;%%
  %4 citations counted in INSPIRE as of 16 juin 2015

%\cite{Chatrchyan:2012ufa}
\bibitem{Chatrchyan:2012ufa} 
  S.~Chatrchyan {\it et al.}  [CMS Collaboration],
  ``Observation of a new boson at a mass of 125 GeV with the CMS experiment at the LHC,''
  Phys.\ Lett.\ B {\bf 716}, 30 (2012)
  [arXiv:1207.7235 [hep-ex]].
  %%CITATION = ARXIV:1207.7235;%%
  %4430 citations counted in INSPIRE as of 22 juin 2015

%\cite{Aad:2012tfa}
\bibitem{Aad:2012tfa} 
  G.~Aad {\it et al.}  [ATLAS Collaboration],
  ``Observation of a new particle in the search for the Standard Model Higgs boson with the ATLAS detector at the LHC,''
  Phys.\ Lett.\ B {\bf 716}, 1 (2012)
  [arXiv:1207.7214 [hep-ex]].
  %%CITATION = ARXIV:1207.7214;%%
  %4514 citations counted in INSPIRE as of 22 juin 2015
  
  %\cite{Wilson:1973jj}
\bibitem{Wilson:1973jj}
  K.~G.~Wilson and J.~B.~Kogut,
  ``The Renormalization group and the epsilon expansion,''
  Phys.\ Rept.\  {\bf 12} (1974) 75.
  %%CITATION = PRPLC,12,75;%%
  %2185 citations counted in INSPIRE as of 27 Jul 2015
  
  %\cite{'tHooft:1979bh}
\bibitem{'tHooft:1979bh}
  G.~'t Hooft,
  ``Naturalness, chiral symmetry, and spontaneous chiral symmetry breaking,''
  NATO Sci.\ Ser.\ B {\bf 59} (1980) 135.
  %176 citations counted in INSPIRE as of 27 Jul 2015
  
  %\cite{Abbott:1984qf}
\bibitem{Abbott:1984qf}
  L.~F.~Abbott,
  ``A Mechanism for Reducing the Value of the Cosmological Constant,''
  Phys.\ Lett.\ B {\bf 150} (1985) 427.
  %%CITATION = PHLTA,B150,427;%%
  %117 citations counted in INSPIRE as of 27 Jul 2015
  
  %\cite{Dvali:2003br}
\bibitem{Dvali:2003br}
  G.~Dvali and A.~Vilenkin,
  ``Cosmic attractors and gauge hierarchy,''
  Phys.\ Rev.\ D {\bf 70} (2004) 063501
  [hep-th/0304043].
  %%CITATION = HEP-TH/0304043;%%
  %21 citations counted in INSPIRE as of 27 Jul 2015
  
  %\cite{Dvali:2004tma}
\bibitem{Dvali:2004tma}
  G.~Dvali,
  ``Large hierarchies from attractor vacua,''
  Phys.\ Rev.\ D {\bf 74} (2006) 025018
  [hep-th/0410286].
  %%CITATION = HEP-TH/0410286;%%
  %22 citations counted in INSPIRE as of 27 Jul 2015

%\cite{Cheung:2007st}
\bibitem{Cheung:2007st} 
  C.~Cheung, P.~Creminelli, A.~L.~Fitzpatrick, J.~Kaplan and L.~Senatore,
  ``The Effective Field Theory of Inflation,''
  JHEP {\bf 0803}, 014 (2008)
  [arXiv:0709.0293 [hep-th]].
  %%CITATION = ARXIV:0709.0293;%%
  %371 citations counted in INSPIRE as of 24 Jun 2015

%\cite{Low:2001bw}
\bibitem{Low:2001bw} 
  I.~Low and A.~V.~Manohar,
  ``Spontaneously broken space-time symmetries and Goldstone's theorem,''
  Phys.\ Rev.\ Lett.\  {\bf 88}, 101602 (2002)
  [hep-th/0110285].
  %%CITATION = HEP-TH/0110285;%%
  %93 citations counted in INSPIRE as of 22 juil. 2015

%\cite{Chluba:2015bqa}
\bibitem{Chluba:2015bqa} 
  J.~Chluba, J.~Hamann and S.~P.~Patil,
  ``Features and New Physical Scales in Primordial Observables: Theory and Observation,''
  IJMPD Vol.\ \textbf{24}, No. 8 1530023 (2015)
  [arXiv:1505.01834 [astro-ph.CO]].
  %%CITATION = ARXIV:1505.01834;%%
  %2 citations counted in INSPIRE as of 24 Jun 2015

%\cite{Dvali:2002pe}
\bibitem{Dvali:2002pe} 
  G.~Dvali, G.~Gabadadze and M.~Shifman,
  ``Diluting cosmological constant in infinite volume extra dimensions,''
  Phys.\ Rev.\ D {\bf 67}, 044020 (2003)
  [hep-th/0202174].
  %%CITATION = HEP-TH/0202174;%%
  %164 citations counted in INSPIRE as of 22 juil. 2015

%\cite{Dvali:2007kt}
\bibitem{Dvali:2007kt} 
  G.~Dvali, S.~Hofmann and J.~Khoury,
  ``Degravitation of the cosmological constant and graviton width,''
  Phys.\ Rev.\ D {\bf 76}, 084006 (2007)
  [hep-th/0703027 [HEP-TH]].
  %%CITATION = HEP-TH/0703027;%%
  %165 citations counted in INSPIRE as of 22 juil. 2015'

%\cite{Aghababaie:2003wz}
\bibitem{Aghababaie:2003wz} 
  Y.~Aghababaie, C.~P.~Burgess, S.~L.~Parameswaran and F.~Quevedo,
  ``Towards a naturally small cosmological constant from branes in 6-D supergravity,''
  Nucl.\ Phys.\ B {\bf 680}, 389 (2004)
  [hep-th/0304256].
  %%CITATION = HEP-TH/0304256;%%
  %262 citations counted in INSPIRE as of 22 juil. 2015
  
%\cite{Burgess:2011va}
\bibitem{Burgess:2011va} 
  C.~P.~Burgess and L.~van Nierop,
  ``Technically Natural Cosmological Constant From Supersymmetric 6D Brane Backreaction,''
  Phys.\ Dark Univ.\  {\bf 2}, 1 (2013)
  [arXiv:1108.0345 [hep-th]].
  %%CITATION = ARXIV:1108.0345;%%
  %21 citations counted in INSPIRE as of 22 juil. 2015
  
%\cite{Spradlin:2001pw}
\bibitem{Spradlin:2001pw} 
  M.~Spradlin, A.~Strominger and A.~Volovich,
  ``Les Houches lectures on de Sitter space,''
  hep-th/0110007.
  %%CITATION = HEP-TH/0110007;%%
  %296 citations counted in INSPIRE as of 14 juil. 2015  

%\cite{Ade:2015xua}
\bibitem{Ade:2015xua} 
  P.~A.~R.~Ade {\it et al.} [Planck Collaboration],
  ``Planck 2015 results. XIII. Cosmological parameters,''
  arXiv:1502.01589 [astro-ph.CO].
  %%CITATION = ARXIV:1502.01589;%%
  %423 citations counted in INSPIRE as of 14 juil. 2015
  
%\cite{Lyth:2001nq}
\bibitem{Lyth:2001nq} 
  D.~H.~Lyth and D.~Wands,
  ``Generating the curvature perturbation without an inflaton,''
  Phys.\ Lett.\ B {\bf 524}, 5 (2002)
  [hep-ph/0110002].
  %%CITATION = HEP-PH/0110002;%%

%\cite{Lyth:2002my}
\bibitem{Lyth:2002my} 
  D.~H.~Lyth, C.~Ungarelli and D.~Wands,
  ``The Primordial density perturbation in the curvaton scenario,''
  Phys.\ Rev.\ D {\bf 67}, 023503 (2003)
  [astro-ph/0208055].
  %%CITATION = ASTRO-PH/0208055;%%
  %540 citations counted in INSPIRE as of 14 juil. 2015

%\cite{Dimopoulos:2003ii}
\bibitem{Dimopoulos:2003ii} 
  K.~Dimopoulos, G.~Lazarides, D.~Lyth and R.~Ruiz de Austri,
  ``The Peccei-Quinn field as curvaton,''
  JHEP {\bf 0305}, 057 (2003)
  [hep-ph/0303154].
  %%CITATION = HEP-PH/0303154;%%
  %73 citations counted in INSPIRE as of 16 Jul 2015

%\cite{Dimopoulos:2003az}
\bibitem{Dimopoulos:2003az} 
  K.~Dimopoulos, D.~H.~Lyth, A.~Notari and A.~Riotto,
  ``The Curvaton as a pseudoNambu-Goldstone boson,''
  JHEP {\bf 0307}, 053 (2003)
  [hep-ph/0304050].
  %%CITATION = HEP-PH/0304050;%%
  %107 citations counted in INSPIRE as of 16 juil. 2015

%\cite{Bozza:2003pj}
\bibitem{Bozza:2003pj} 
  V.~Bozza, M.~Gasperini, M.~Giovannini and G.~Veneziano,
  ``The axion as a curvaton in string cosmology models,''
  %%CITATION = INSPIRE-640066;%%
%\cite{Miele:2004ic}
in   G.~Miele and G.~Longo,
  ``Thinking, observing and mining the universe. Proceedings, International Conference, Sorrento, Italy, September 22-27, 2003,''
  %%CITATION = INSPIRE-672206;%%

%\cite{Mazumdar:2015pta}
\bibitem{Mazumdar:2015pta} 
  A.~Mazumdar and S.~Qutub,
  ``Non-perturbative over-production of axion-like-particles (ALPs) via derivative interaction,''
  arXiv:1508.04136 [hep-ph].
  %%CITATION = ARXIV:1508.04136;%%
  %1 citations counted in INSPIRE as of 20 Dec 2015

  %\cite{Espinosa:2015eda}
\bibitem{Espinosa:2015eda}
  J.~R.~Espinosa, C.~Grojean, G.~Panico, A.~Pomarol, O.~Pujolàs and G.~Servant,
  ``Cosmological Higgs-Axion Interplay for a Naturally Small Electroweak Scale,''
  arXiv:1506.09217 [hep-ph].
  %%CITATION = ARXIV:1506.09217;%%
  
  %\cite{Hardy:2015laa}
\bibitem{Hardy:2015laa}
  E.~Hardy,
  ``Electroweak relaxation from finite temperature,''
  arXiv:1507.07525 [hep-ph].
  %%CITATION = ARXIV:1507.07525;%%
  
  %\cite{Bai:2013xga}
\bibitem{Bai:2013xga}
  Y.~Bai and P.~Schwaller,
  ``Scale of dark QCD,''
  Phys.\ Rev.\ D {\bf 89} (2014) 6,  063522
  [arXiv:1306.4676 [hep-ph]].
  %%CITATION = ARXIV:1306.4676;%%
  %22 citations counted in INSPIRE as of 29 Jul 2015
  
  %\cite{Juknevich:2009gg}
\bibitem{Juknevich:2009gg}
  J.~E.~Juknevich,
  ``Pure-glue hidden valleys through the Higgs portal,''
  JHEP {\bf 1008} (2010) 121
  [arXiv:0911.5616 [hep-ph]].
  %%CITATION = ARXIV:0911.5616;%%
  %12 citations counted in INSPIRE as of 29 juil. 2015
  
  %\cite{Craig:2015pha}
\bibitem{Craig:2015pha}
  N.~Craig, A.~Katz, M.~Strassler and R.~Sundrum,
  ``Naturalness in the Dark at the LHC,''
  JHEP {\bf 1507} (2015) 105
  [arXiv:1501.05310 [hep-ph]].
  %%CITATION = ARXIV:1501.05310;%%
  %15 citations counted in INSPIRE as of 29 Jul 2015
  
  %\cite{Curtin:2015fna}
\bibitem{Curtin:2015fna}
  D.~Curtin and C.~B.~Verhaaren,
  ``Discovering Uncolored Naturalness in Exotic Higgs Decays,''
  arXiv:1506.06141 [hep-ph].
  %%CITATION = ARXIV:1506.06141;%%
  
  %\cite{Schwaller:2015gea}
\bibitem{Schwaller:2015gea}
  P.~Schwaller, D.~Stolarski and A.~Weiler,
  ``Emerging Jets,''
  JHEP {\bf 1505} (2015) 059
  [arXiv:1502.05409 [hep-ph]].
  %%CITATION = ARXIV:1502.05409;%%
  %14 citations counted in INSPIRE as of 29 Jul 2015
  
  %\cite{Strassler:2006im}
\bibitem{Strassler:2006im}
  M.~J.~Strassler and K.~M.~Zurek,
  ``Echoes of a hidden valley at hadron colliders,''
  Phys.\ Lett.\ B {\bf 651} (2007) 374
  [hep-ph/0604261].
  %%CITATION = HEP-PH/0604261;%%
  %356 citations counted in INSPIRE as of 29 Jul 2015
  
  %\cite{GarciaBellido:1999sv}
\bibitem{GarciaBellido:1999sv}
  J.~Garcia-Bellido, D.~Y.~Grigoriev, A.~Kusenko and M.~E.~Shaposhnikov,
  %``Nonequilibrium electroweak baryogenesis from preheating after inflation,''
  Phys.\ Rev.\ D {\bf 60} (1999) 123504
  [hep-ph/9902449].
  %%CITATION = HEP-PH/9902449;%%
  %172 citations counted in INSPIRE as of 29 juil. 2015
  
  %\cite{Krauss:1999ng}
\bibitem{Krauss:1999ng}
  L.~M.~Krauss and M.~Trodden,
  %``Baryogenesis below the electroweak scale,''
  Phys.\ Rev.\ Lett.\  {\bf 83} (1999) 1502
  [hep-ph/9902420].
  %%CITATION = HEP-PH/9902420;%%
  %102 citations counted in INSPIRE as of 29 juil. 2015
  
  %\cite{Tranberg:2003gi}
\bibitem{Tranberg:2003gi}
  A.~Tranberg and J.~Smit,
  %``Baryon asymmetry from electroweak tachyonic preheating,''
  JHEP {\bf 0311} (2003) 016
  [hep-ph/0310342].
  %%CITATION = HEP-PH/0310342;%%
  %75 citations counted in INSPIRE as of 29 juil. 2015
  
  %\cite{Servant:2014bla}
\bibitem{Servant:2014bla}
  G.~Servant,
  %``Baryogenesis from Strong $CP$ Violation and the QCD Axion,''
  Phys.\ Rev.\ Lett.\  {\bf 113} (2014) 17,  171803
  [arXiv:1407.0030 [hep-ph]].
  %%CITATION = ARXIV:1407.0030;%%
  %3 citations counted in INSPIRE as of 29 juil. 2015

\bibitem{Toni}  
 A.~Kehagias, T.~Riotto \textit{private communication}
\end{thebibliography}
\end{document}